\documentclass[aps,twocolumn,showpacs,prb]{revtex4}

\usepackage{color,vmargin,epsfig,rotating,graphicx,subfigure}
\usepackage{amsmath,amssymb,amscd}
\usepackage[latin1]{inputenc}
\usepackage[english]{babel}
\usepackage{dsfont}
\usepackage{textcomp}

\setpapersize{A4}

\renewcommand{\Im}{{\rm Im}}

\newcommand{\ri}{{\rm i}}
\newcommand{\re}{{\rm e}}
\newcommand{\rd}{{\rm d}}

\newcommand{\Tr}{{\rm Tr}}

\newcommand{\reff}[1]{(\ref{#1})}
\newcommand{\vk}{\vec{\kappa}}
\newcommand{\vrh}{\vec{\rho}}
\newcommand{\vj}{\vec{j}}
\newcommand{\va}{\vec{a}}
\newcommand{\vq}{\vec{q}}
\newcommand{\skp}{(\vk \cdot \vk')}
\newcommand{\skps}{(\vk' \cdot \vk'')}
\newcommand{\kp}{(\vk \times \vk')_z}
\newcommand{\kps}{(\vk' \times \vk'')_z}
\newcommand{\F}[1]{\widehat{F}^{(#1)}}
\newcommand{\ts}{t_{\rm s}(\vk)}
\newcommand{\tp}{t_{\rm p}(\vk)}
\newcommand{\tss}{t_{\rm s}(\vk')}
\newcommand{\tps}{t_{\rm p}(\vk')}
\newcommand{\tsss}{t_{\rm s}(\vk'')}
\newcommand{\tpss}{t_{\rm p}(\vk'')}

\renewcommand{\vec}{\boldsymbol}

\begin{document}

\title[LDOS above a nanostructured surface] {Second-order calculation
	of the local density of states above a nanostructured surface} 

\author{Felix R\"uting}
\author{Svend-Age Biehs}
\altaffiliation[Present address: ]{Laboratoire Charles Fabry, 
	Institut d'Optique, CNRS, Universit\'e Paris-Sud, 
	Campus Polytechnique, RD128, 91127 Palaiseau cedex, France}
\author{Oliver Huth}
\author{Martin Holthaus}

\affiliation{Institut f\"ur Physik, Carl von Ossietzky Universit\"at,
    	D-26111 Oldenburg, Germany}

\date{21th April 2010}

\begin{abstract}
We have numerically implemented a perturbation series for the scattered 
electromagnetic fields above rough surfaces, due to Greffet, allowing us 
to evaluate the local density of states to second order in the surface
profile function. We present typical results for thermal near fields of
surfaces with regular nanostructures, investigating the relative magnitude 
of the contributions appearing in successive orders. The method is then 
employed for estimating the resolution limit of an idealized Near-Field 
Scanning Thermal Microscope (NSThM). 
\end{abstract}




\maketitle


\section{Introduction}
\label{Sec:S_1}

Quite recently there has been a notable increase of experimental activities 
aiming at the exploration of properties of thermally generated fluctuating 
electromagnetic fields close to the surface of some material, and at detecting 
the near-field mediated heat transfer.~\cite{RytovEtAl89,JoulainEtAl05,
VolokitinPersson07} Hu {\em et al.\/} have measured the near-field thermal 
radiation between \textmu m-spaced glass plates, and have demonstrated that 
the resulting near-field heat transfer exceeds the far-field limit set by 
Planck's blackbody radiation law.~\cite{HuEtAl08} Next, Narayanaswamy 
{\em et al.\/} and Shen {\em et al.\/} have studied the heat transfer between 
microspheres and flat substrates, with emphasis on the coupling of surface
phonon polaritons across the gap between them, and have reported heat 
transfer coefficients three orders of magnitude above the blackbody radiation
limit.~\cite{NarayanaswamyEtAl08,ShenEtAl09} Then Rousseau {\em et al.\/}
have carried out precise measurements of the radiative heat transfer between 
sodalime glass spheres with diameters of 22 or 40 \textmu m and 
borosilicate glass plates for distances ranging from 30~nm to 
2.5~\textmu m,~\cite{RousseauEtAl09} and have verified theoretical predictions 
based on fluctuational electrodynamics~\cite{RytovEtAl89,PolderVanHove71} with 
impressive accuracy. On the other hand, significant progress has been made at 
using near-field effects for imaging. Kittel {\em et al\/.} are developing 
a device termed Near-Field Scanning Thermal Microscope 
(NSThM)~\cite{MuellerHirschEtAl99,KittelEtAl05,WischnathEtAl08} which does not
yet seem capable of highly accurate quantitative measurements of the near-field 
heat current between its sensor and the sample, but which lends itself to 
nanoscale thermal imaging of structured surfaces.~\cite{KittelEtAl08} Moreover,
De Wilde {\em et al.\/} have reported the successful operation of a Thermal 
Radiation Scanning Tunneling Microscope (TRSTM),~\cite{DeWildeEtAl06} providing
images of thermally excited surface plasmons, and giving clear evidence for 
spatial coherence effects in near-field thermal emission. 

These remarkable developments indicate that thermal near-field physics, 
after having been under intense theoretical investigation for some time 
already,~\cite{RytovEtAl89,JoulainEtAl05,VolokitinPersson07,PolderVanHove71}
is breaking through to the forefront of experimental research right now.
There are several compelling reasons for this trend: Besides the prospects of 
obtaining novel insight into fundamental physics in dielectric matter, and of 
developing advanced diagnostic tools for materials science, thermal near-field 
effects have great potential for near-field thermo\-photo\-voltaic energy 
conversion.~\cite{DiMatteoEtAl01,NarayanaswamyChen03,LarocheEtAl06,
FrancoeurEtAl08,BasuEtAl09}   
     
On the theoretical side, one of the most important quantities
characterizing the fluctuating thermal near field close to a
dielectric surface is its local density of states
(LDOS).~\cite{JoulainEtAl03} In particular, the power~$P$ transferred
between a dielectric sample at temperature~$T_{\rm S}$ and a nanoparticle at
temperature~$T_{\rm P}$, brought into the sample's near field at a position
$\vec{a}$ such that the particle may effectively be treated within the
dipole approximation, and the heat transfer proceeds almost entirely
via evanescent modes, is given by (see, e.g.,
Refs.~\onlinecite{Dorofeyev98,Pendry99,MuletEtAl01,DedkovKyasov07,
  Dorofeyev08,ChapuisEtAl08})
\begin{align}
\label{eq:Pdip}
	P = & \int\limits_0^{\infty} \! \rd \omega \, 
	2 \omega \left[ \Theta(\omega, T_{\rm P}) - \Theta(\omega, T_{\rm S}) \right] \\
	& \times \left[\alpha_{\rm P}''(\omega) D^{\rm E}(\omega,\vec{a}) +
  		\mu_{\rm P}''(\omega) D^{\rm H}(\omega,\vec{a}) \right] \; ,
\nonumber
\end{align}
where $D^{\rm E}(\omega,\vec{a})$ is the electric and $D^{\rm H}(\omega,\vec{a})$ is
the magnetic part of the sample's LDOS at the point $\vec{a}$ of effective 
interaction; $\alpha_{\rm P}''(\omega)$ and $\mu_{\rm P}''(\omega)$ denote the 
imaginary part of the particle's electric and magnetic polarizability,
respectively. Finally,
\begin{equation}
	\Theta(\omega, T) = \frac{\hbar\omega}
	{\exp(\hbar\omega/k_{\rm B}T) - 1}	
\end{equation}
is the Bose-Einstein function; the sign in Eq.~(\ref{eq:Pdip}) is chosen such 
that a net energy transfer from the particle to the sample, occurring for 
$T_{\rm P} > T_{\rm S}$, gives a positive~$P$. The use of the dipole 
approximation underlying Eq.~(\ref{eq:Pdip}) requires that the distance of
the nanoparticle from the sample's surface remains large compared to its linear
size. Those frequencies which significantly contribute to the heat transfer 
are limited by the higher temperature $T_{\max}=\max(T_{\rm P},T_{\rm S})$. 
Provided the polarizabilities and the LDOS exhibit no resonances in the 
accessible frequency regime, requiring in particular the absence of thermally 
excitable surface modes, the main contribution to the integral~\reff{eq:Pdip} 
merely stems from frequencies in the vicinity of the thermal frequency 
$\omega_{\rm th} \approx 2.82 \; k_{\rm B}T/\hbar$, so that 
\begin{align}
\label{eq:dipapprox}
	P \propto \omega_{\rm th} & 
	\big[\alpha_{\rm P}''(\omega_{\rm th}) D^{\rm E}(\omega_{\rm th},\vec{a}) \\
        & + \mu_{\rm P}''(\omega_{\rm th}) D^{\rm H}(\omega_{\rm th},\vec{a}) \big] \; .
\nonumber
\end{align}
Under suitable conditions, already this simple approximation can give 
surprisingly good results when trying to theoretically reconstruct
surface images obtained with the NSThM.~\cite{KittelEtAl08}

The possibility to experimentally assess the LDOS above nanostructured 
surfaces demands refined techniques for its calculation. Assuming local 
thermal equilibrium, and considering positions~$\vec{r}$ so close to the 
sample's surface that the energy density is dominated by evanescent modes 
and the contribution of propagating modes can be neglected, the electric 
and the magnetic LDOS are related to the imaginary parts of the traces of 
the renormalized (or `reflected') electric and magnetic Green's dyadics 
$\mathds{G}^{\rm E}_{\rm r}$ and $\mathds{G}^{\rm H}_{\rm r}$ through the 
relations~\cite{JoulainEtAl03}
\begin{equation}
	D^{\rm E}(\omega,\vec{r}) = \frac{\omega}{\pi c^2} 
	\Im \, \Tr \, \mathds{G}^{\rm E}_{\rm r}(\vec{r},\vec{r})
\end{equation}
and
\begin{equation}
	D^{\rm H}(\omega,\vec{r}) = \frac{\omega}{\pi c^2} 
	\Im \, \Tr \, \mathds{G}^{\rm H}_{\rm r}(\vec{r},\vec{r}) \; .
\end{equation}
In the present paper we exploit this connection for computing the LDOS above 
a nano\-structured surface to second order in the surface profile, relying on 
an earlier formulation of the perturbation series by Greffet.~\cite{Greffet88}
The method is technically involved, and soon hits practical computational
limits when proceeding to higher orders. Nonetheless, we show that 
second-order calculations now are feasible routinely. Our work thus extends
a previous study,~\cite{BiehsEtAl08} which has given first-order results, 
and enables us to delineate under which conditions low-order perturbation
theory is sufficient. It also complements a recent investigation by Biehs 
and Greffet~\cite{BiehsGreffet09} who have considered rough surfaces 
described as stochastic Gaussian processes. In contrast, we focus on surfaces 
with deliberately induced, regular nanoprofiles. We proceed as follows: 
In Sec.~\ref{Sec:S_2} we sketch the underlying perturbative 
scheme,~\cite{Greffet88} and outline a few details of its numerical 
implementation, deferring technicalities to the Appendix~\ref{App:A}. We then 
present results of our computations in Sec.~\ref{Sec:S_3}, first examining the 
relative magnitude of first-and second-order contributions, and then outlining 
how to quantify the resolution power of an idealized NSThM. Some conclusions 
are drawn in the final Sec.~\ref{Sec:S_4}.

\section{Computation of the Green's dyadics}
\label{Sec:S_2}

In this section we utilize an analytical perturbative approach, originally 
developed by Greffet for calculating the scattered electromagnetic waves above 
a rough dielectric surface,~\cite{Greffet88} in order to obtain the required
Green's dyadics $\mathds{G}^{\rm E}_{\rm r}$ and $\mathds{G}^{\rm H}_{\rm r}$. Greffet's approach
results in a series of recursively determined contributions in ascending 
orders of the surface profile, and thus enables one to systematically assess 
the higher-order terms.

\subsection{Calculational scheme}    

We assume that the surface is described by an expression $z = hf(x,y)$,
where $f(x,y)$ is a normalized function varying between $+1$ and $-1$; the 
scaling parameter~$h$ carries the dimension of a length. The nonmagnetic 
dielectric medium, equipped with permittivity $\epsilon(\omega)$, fills the 
entire half-space $z<hf(x,y)$. For $z>hf(x,y)$, outside the dielectric, 
the total electric field consists of a prescribed incident component 
$\vec{E}_{\rm i}(\vec{r})$, and of the so far unknown reflected component 
$\vec{E}_{\rm r}(\vec{r})$, while the transmitted field inside the dielectric is 
denoted as $\vec{E}_{\rm t}(\vec{r})$. Greffet has given a recursive series 
solution for the transmitted and the reflected field,~\cite{Greffet88} invoking
the extinction theorem and the Rayleigh hypothesis.~\cite{FariasMaradudin83} 
The extinction theorem amounts to an exact integral formulation of the boundary
condition, whereas the use of the Rayleigh hypothesis means expanding the 
transmitted, incident, and reflected fields in plane waves travelling in 
$z$-direction,
\begin{equation}
\label{eq:exp_et}
	\vec{E}_{\rm t}(\vec{r}) = \int \! \rd^2 \kappa \,
	\vec{e}_{\rm t}(\vk)\re^{\ri(\vk\cdot\vec{\rho}-k_z z)} \; ,
\end{equation}
\begin{equation}
	\vec{E}_{\rm i}(\vec{r}) = \int \! \rd^2 \kappa \,
	\vec{e}_{\rm i}(\vk)\re^{\ri(\vk\cdot\vec{\rho}-k_{z0} z)} \; ,
\end{equation}
and
\begin{equation}
\label{eq:exp_er}
	\vec{E}_{\rm r}(\vec{r}) = \int \! \rd^2 \kappa \,
	\vec{e}_{\rm r}(\vk)\re^{\ri(\vk\cdot\vec{\rho}+k_{z0} z)} \; .
\end{equation}
Here we write $\vec{r}=[x,y,z]^t$ for the position vector,
$\vec{\rho}=[x,y,0]^t$ for its lateral part, and $\vk=[k_x,k_y,0]^t$ for
the lateral component of the wave vector; moreover,
\begin{equation}
	k_z=\sqrt{\epsilon k_0- \kappa}
\end{equation}
and 
\begin{equation}
	k_{z0}=\sqrt{k_0-\kappa}
\end{equation}
are the $z$-components of the wave vector inside and outside the medium. We 
also use the notation $k_0 = \omega/c$ and $\kappa=|\vk|$. These expansions 
\reff{eq:exp_et}-\reff{eq:exp_er} assume translational symmetry in the 
$x$-$y$-plane and hence are strictly justified outside the structured region, 
that is, for $z > h$ and $z < -h$. However, the extinction theorem requires 
to evaluate the fields on the very surface of the dielectric, where the
validity of the above expressions cannot be taken for granted. Ignoring 
this complication and using the expansions \reff{eq:exp_et}-\reff{eq:exp_er}  
nonetheless is a common procedure~\cite{HenkelSandoghdar98,LambrechtEtAl06} 
which has been looked into by several authors from the mathematical point of 
view;~\cite{BergFokkema79,HugoninEtAl81,Keller00,Ramm02} it appears to work 
reliably at least for sufficiently small values of~$h$. For example, in the
case of a sinusoidal grating described by $z = (h/2) \cos(2\pi x/D)$ this 
Rayleigh hypothesis holds rigorously~\cite{BergFokkema79,HugoninEtAl81,Keller00}
up to $h_{\max}/D = 0.142\,521$, and may therefore be employed for both the 
propagating and the evanescent parts of the field as long as the ratio $h/D$ 
stays below this boundary. 
 
The field's Fourier components then are expanded in the forms
\begin{equation}
	\vec{e}_{\rm t}(\vk)=\sum\limits_{m=0}^\infty
	\frac{\vec{e}^{(m)}_{\rm t}(\vk)}{m!} \; ,
\end{equation}
\begin{equation}
	\vec{e}_{\rm r}(\vk)=\sum\limits_{m=0}^\infty
	\frac{\vec{e}^{(m)}_{\rm r}(\vk)}{m!} \; ,
\end{equation}
and
\begin{equation}
	\vec{e}_{\rm i}(\vk)=\sum\limits_{m=0}^\infty
	\frac{\vec{e}^{(m)}_{\rm i}(\vk)}{m!} = \vec{e}_{\rm i}^{(0)}(\vk) \; .
\end{equation}
It is useful to split the fields into their s\,- and p\,- components
according to
\begin{equation}
	\vec{e}_{\rm t}(\vec{\kappa})=
	e_{\rm t,s}(\vk)\vec{a}_{\rm s}(\vk)+e_{\rm t,p}(\vk)\vec{a}_{\rm p,t}^-(\vk) \; ,
\end{equation}
\begin{equation}
	\vec{e}_{\rm i}(\vec{\kappa})=
	e_{\rm i,s}(\vk)\vec{a}_{\rm s}(\vk)+e_{\rm i,p}(\vk)\vec{a}_{\rm p,0}^-(\vk) \; ,
\end{equation}
and
\begin{equation}
	\vec{e}_{\rm r}(\vec{\kappa})=
	e_{\rm r,s}(\vk)\vec{a}_{\rm s}(\vk)+e_{\rm r,p}(\vk)\vec{a}_{\rm p,0}^+(\vk) \; , 
\end{equation}
where
\begin{equation}
	\vec{a}_{\rm s}(\vk)=\frac{1}{\kappa}\left[ \begin{array}{c} 
	-k_y \\ k_x \\ 0 \end{array} \right] \; ,
\end{equation}
\begin{equation}
	\vec{a}_{\rm p,t}^-(\vk)=-\frac{1}{n k_0 \kappa}\left[ \begin{array}{c} 
	k_x k_z \\ k_y k_z \\ \kappa^2 \end{array} \right] \; ,
\end{equation}
\begin{equation}
	\vec{a}_{\rm p,0}^-(\vk)=-\frac{1}{k_0 \kappa}\left[ \begin{array}{c} 
	k_x k_{z0} \\ k_y k_{z0} \\ \kappa^2 \end{array} \right] \; ,
\end{equation}
and
\begin{equation}
	\vec{a}_{\rm p,0}^+(\vk)=\frac{1}{k_0 \kappa}\left[ \begin{array}{c} 
	k_x k_{z0} \\ k_y k_{z0} \\ -\kappa^2 \end{array} \right] \; ;
\end{equation}
here $n=\sqrt{\epsilon}$ is the index of refraction. 

Following Greffet,~\cite{Greffet88} one then obtains the transmitted field 
in the recursive form
\begin{align}
\label{eq:rec_et}
	\left[\begin{array}{c} 
	e_{\rm t,s}^{(m)}(\vk) \\ e_{\rm t,p}^{(m)}(\vk) \end{array}\right] & =
	\frac{k_z-k_{z0}}{4 \pi^2}\mathds{R}^{-1}(\vk,\vk)
	\int \! \rd^2 \kappa' \, \Bigg\{\mathds{R}(\vk,\vk')
\nonumber\\
	& \times \sum\limits_{q=1}^{m} \binom{m}{q}\frac{(\ri h)^q
	\F{q}(\vk'-\vk)}{(k_{z0}-k_z')^{1-q}}
\nonumber\\
 	&\times\left[ \begin{array}{c} 
	e_{\rm t,s}^{(m-q)}(\vk') \\ e_{\rm t,p}^{(m-q)}(\vk') \end{array} \right]
	\Bigg\} \; ,
\end{align}
so that $\vec{e}^{(m)}_{\rm t}(\vk)$ is proportional to $h^m$; the according 
expression for the reflected field reads
\begin{align}
\label{eq:rec_er}
	\left[\begin{array}{c} 
	e_{\rm r,s}^{(m)}(\vk) \\ e_{\rm r,p}^{(m)}(\vk) \end{array}\right] & =
	\frac{\epsilon-1}{8 \pi^2 k_{z0}} \Bigg\{ 
	\int \! \rd^2 \kappa' \, \Bigg( \mathds{P}(\vk,\vk')
\nonumber\\
	&\hspace{-1 cm}\times \sum\limits_{q=0}^{m-1}\binom{m}{q}
	\frac{(-\ri h)^{m-q}\F{m-q}(\vk'-\vk)}{(k_{z0}+k_z')^{1+q-m}}
\nonumber\\
	&\hspace{-1 cm}\times \left[
    	\begin{array}{c} 
	e_{\rm t,s}^{(q)}(\vk') \\ e_{\rm t,p}^{(q)}(\vk') \end{array} \right]\Bigg)
	\nonumber\\
	&\hspace{-1cm}+\mathds{P}(\vk,\vk)\frac{4 \pi^2}{k_z+k_{z0}}
	\left[\begin{array}{c} 
	e_{\rm t,s}^{(m)}(\vk) \\ e_{\rm t,p}^{(m)}(\vk) \end{array}
    	\right]\Bigg\} \; .
\end{align}
In these equations the quantities $\widehat{F}^{(n)}(\vk)$ denote the Fourier 
transforms of powers of the surface function,  
\begin{equation}
	\widehat{F}^{(n)}(\vk)=\int \! \rd^2 \rho \, 
	\re^{\ri \vk \cdot \vec{\rho}} f^n(\vec{\rho}) \; .
\end{equation}
The linear operators $\mathds{R}(\vk,\vk')$ and $\mathds{P}(\vk,\vk')$ 
effectuate the double vectorial product with $\vec{k}_{\rm r}^-=[k_x, k_y,-k_{z0}]^t$ and 
$\vec{k}_{\rm r}^+=[k_x, k_y, k_{z0}]^t$, respectively; these double products (namely, 
$\vec{k}_{\rm r}^- \times \vec{k}_{\rm r}^- \times$ and $\vec{k}_{\rm r}^+ \times \vec{k}_{\rm r}^+ \times$) typically 
appear when using the extinction theorem. In the basis chosen, the matrix
forms of these operators are 
\begin{equation}
	\mathds{R}(\vk,\vk')=-\frac{k_0^2}{\kappa \kappa'}
	\left[ \begin{array}{c c} \vk \cdot \vk' &
		-\frac{k'_z(\vk \times \vk')_z}{n  k_0} \\
		\frac{k_{z0}(\vk \times \vk')_z}{k_0} & 
		\frac{\kappa^2 \kappa'^2 + \vk \cdot \vk' k_{z0}k'_z}{n k_0^2}
	\end{array}\right]
\end{equation}
and
\begin{equation}
	\mathds{P}(\vk,\vk')=-\frac{k_0^2}{\kappa \kappa'}
	\left[ \begin{array}{c c} \vk \cdot \vk' &
		-\frac{k'_z(\vk \times \vk')_z}{n  k_0} \\
		\frac{-k_{z0}(\vk \times \vk')_z}{k_0} & 
		\frac{\kappa^2 \kappa'^2 - \vk \cdot \vk' k_{z0}k'_z}{n k_0^2}
	\end{array}\right] \; ,
\end{equation}
writing $(\vk \times \vk')_z$ for the $z$-component of the vectorial 
product of $\vk$ and $\vk'$.

The above recursions start from the well known half-space results obtained
for a perfectly flat surface, which can be cast into the forms 
\begin{equation}
	\left[ \begin{array}{c} 
	e_{\rm t,s}^{(0)}(\vk) \\ e_{\rm t,p}^{(0)}(\vk) 
	\end{array} \right ] = \left[ \begin{array}{cc} 
	t_{\rm s}(\vk) & 0 \\ 0 & t_{\rm p}(\vk) 
	\end{array} \right] \left[ \begin{array}{c} 
	e_{\rm i,s}(\vk) \\ e_{\rm i,p}(\vk) 
	\end{array} \right ]
\end{equation}
and
\begin{equation}
	\left[ \begin{array}{c} 
	e_{\rm r,s}^{(0)}(\vk) \\ e_{\rm r,p}^{(0)}(\vk) 
	\end{array} \right ] = \left[ \begin{array}{cc}  
	r_{\rm s}(\vk) & 0 \\ 0 & r_{\rm p}(\vk) 
	\end{array} \right] \left[ \begin{array}{c}
  	e_{\rm i,s}(\vk) \\ e_{\rm i,p}(\vk) 
	\end{array} \right ] 
\end{equation}
with the Fresnel coefficients
\begin{eqnarray*}
	t_{\rm s}(\vk) = \frac{2 k_{z0}}{k_{z0}+k_z}		& ; & 
	t_{\rm p}(\vk) = \frac{2 n k_{z0}}{n^2 k_{z0}+k_z} \; ;
\\\	
	r_{\rm s}(\vk) = \frac{k_{z0}-k_z}{k_{z0}+k_z}	& ; &
	r_{\rm p}(\vk) = \frac{n^2k_{z0}-k_z }{n^2k_{z0}+k_z} \; . 
\end{eqnarray*}

From these fields~\reff{eq:rec_et}-\reff{eq:rec_er} we now proceed to the 
calculation of the Green's dyadics. More precisely, in order to compute the
local density of states we need to determine the `reflected' part of the 
Green's dyadics for coinciding source and observation points. To this end, we 
take the field of a delta-like source current located at $\vrh+ z\vec{e}_z$
and pointing into the direction of the unit vector $\vj$ as incident field,
giving
\begin{equation}
	\left[ \begin{array}{c}
	e_{\rm i,s}(\vk) \\ e_{\rm i,p}(\vk)
       \end{array} \right] =
       -\frac{\omega \mu_0}{2 k_{z0}}\re^{\ri(\vk \cdot \vrh+k_{z0} z)}
	\left[ \begin{array}{c}
	\va_{\rm s}(\vk)\cdot \vj \\ \va^-_{\rm p,0}(\vk) \cdot \vj
	\end{array} \right] \; . 
\end{equation}
To zeroth order in $h$, the reflected field then is
\begin{align}
\label{eq:Eref0}
	\left[ \begin{array}{c}
	e_{\rm r,s}^{(0)}(\vk) \\ e_{\rm r,p}^{(0)}(\vk)
	\end{array} \right] & =
	-\frac{\omega \mu_0}{2 k_{z0}}\re^{\ri(\vk \cdot \vrh+k_{z0} z)} \\
	&\times \left[ \begin{array}{cc}
	r_{\rm s}(\vk) & 0 \\ 0 & r_{\rm p}(\vk)
	\end{array} \right] \left[ \begin{array}{c}
	\va_{\rm s}(\vk)\cdot \vj \\ \va^+_{\rm p,0}(\vk) \cdot \vj
	\end{array} \right] \; .
\nonumber
\end{align}
The directions of the s\,- and p\,-components of the incident field are given 
by $\va_{\rm s}(\vk)$ and $\va^-_{\rm p,0}(\vk)$, while the components of the 
reflected field are given by $\va_{\rm s}(\vk)$ and $\va^+_{\rm p,0}(\vk)$. 
Therefore it is convenient to split the Fourier coefficients of the Green's 
dyadics into the four parts that result from taking the dyadic products of 
these unit vectors, leading to
\begin{align}
	g_{\rm r}^{\rm E,(0)}(\vk)&=g^{\rm E,(0)}_{\rm r,ss}(\vk)\va_{\rm s}(\vk)\otimes \va_{\rm s}(\vk) \\
	&+g^{\rm E,(0)}_{\rm r,sp}(\vk)\va_{\rm s}(\vk)\otimes \va^-_{\rm p,0}(\vk)  \nonumber\\
	&+g^{\rm E,(0)}_{\rm r,ps}(\vk)\va^+_{\rm p,0}(\vk)\otimes\va_{\rm s}(\vk)   \nonumber\\
	&+g^{\rm E,(0)}_{\rm r,pp}(\vk)\va^+_{\rm p,0}(\vk)\otimes \va^-_{\rm p,0}(\vk) \; .
\nonumber
\end{align}
Having employed a delta-like source current the relation between the field 
and the electric Green's dyadic simply reads 
$\mathds{G}^{\rm E}\cdot \vj = \vec{E} / (\ri \omega \mu_0)$, so that the 
coefficients of this dyadic can easily be read off from Eq.~\reff{eq:Eref0}. 
By means of an inverse Fourier transform, equating source and observation 
point, one then arrives at the familiar result for the reflected Green's 
dyadic pertaining to a flat surface,
\begin{align}
\label{eq:green0}
	\mathds{G}_{\rm r}^{\rm E,(0)} & = \frac{\ri}{4 \pi^2}\int \! \rd^2\kappa \,
	\frac{\re^{2 \ri k_{z0} z }}{2 k_{z0}}\big[r_{\rm s}(\vk) \va_{\rm s}(\vk) 
	\otimes \va_{\rm s}(\vk) 
\nonumber\\
	&+ r_{\rm p}(\vk) \va^+_{\rm p,0}(\vk) \otimes \va^-_{\rm p,0}(\vk) \big] \; .
\end{align}
With the help of Eq.~\reff{eq:rec_er} one obtains similar expressions to all 
orders in the profile height~$h$. An advantage of this approach lies in the 
fact that it is then quite easy to also determine the magnetic Green's dyadic, 
which is related to the electric one through~\cite{FelsenMarcuvitz94}
\begin{equation}
	\mathds{G}_{\rm r}^{\rm H}(\vec{r},\vec{r}) = -\frac{1}{k_0^2}\nabla \times
	\mathds{G}_{\rm r}^{\rm E}(\vec{r},\vec{r}') \times \nabla' |_{\vec{r}'=\vec{r}} 
	\; .
\end{equation}
In Fourier space the operator $\nabla$ is replaced either by 
$\ri(\vk + k_{z0} \vec{e}_z)$ or by $\ri(\vk - k_{z0} \vec{e}_z)$, depending 
on whether the curl acts on a unit vector belonging to the incident or to the 
reflected field. Therefore, the magnetic Green's dyadic is derived from its 
electric counterpart by simply replacing $\va(\vk) \otimes \va(\vk')$ by the 
expression 
$\frac{1}{k_0^2}(\vk+ k_{z0}\vec{e}_z) \times \va(\vk) \otimes \va(\vk') 
\times (\vk'- k'_{z0}\vec{e}_z)$. Here we introduce $\vk'$ and $k'_{z0}$ 
because it is only to zeroth order that the reflected and the incident field 
are characterized by the same wave vector.

Finally, for calculating the trace of the Green's dyadics one just has to 
replace the dyadic products by their traces. Hence, the method sketched here 
yields a transparent strategy for obtaining the electric and the magnetic 
LDOS above a structured surface. In practice, a restriction on the maximum 
order achievable is imposed by the available computational resources: To 
zeroth order only two-dimensional integrals have to be evaluated, to first 
order four-dimensional integrals appear; to second order one already has
to deal with six-dimensional integrals, and so on. Clearly, a good choice
of the numerical tools is decisive here; we therefore present some details 
of our implementation.

\subsection{Numerical implementation}
\label{Subsec:IIB}

We exemplarily discuss the calculation of the electric local density of states 
to first and second order; its magnetic counterpart, and the higher-order terms,
are determined in a similar way. With the help of Eq.~\reff{eq:rec_er} the 
trace of the first-order contribution to the electric reflected Green's dyadic 
is computed as
\begin{equation}
\label{eq:trace1}
	\Tr\big\{\mathds{G}_{\rm r}^{\rm E,(1)}\big\} = 
	-\int\! \frac{\rd^2 q}{ 4 \pi^2} \,
	\re^{\ri \vq \cdot \vrh} \widehat{F}^{(1)}(-\vq) a_1(\vq) \; ,
\end{equation}
where $a_1(\vq)$ is given by the integral
\begin{equation}
\label{eq:S1}
	a_1(\vq)=\int \! \rd^2 \kappa' \, \vec{S}^{(1)}(\vq+\vk',\vk')\cdot
	\vec{A}^{\rm (E)}_{\rm tr}(\vq+\vk',\vk')
\end{equation}
with the four-dimensional vectors $\vec{S}^{(1)}$ and $\vec{A}^{\rm (E)}_{\rm tr}$ 
specified in Appendix~\ref{App:A}. We either employ an experimentally 
determined surface profile,~\cite{KittelEtAl08} or some suitably selected
model function $f(\vec{\rho})$; sample it, and perform a discrete FFT in order 
to determine $\F{1}$. Then we compute $a(\vq)$ for each required $\vq$ by
numerical integration, and finally take an inverse FFT of $\F{1}(-\vq)a(\vq)$ 
to get the trace~(\ref{eq:trace1}) of the first-order Green's dyadic.

To second order one has to deal with two contributions, one containing $\F{2}$,
the other feeding from two factors $\F{1}$. The former contribution has the 
same structure as the first-order term, 
\begin{align}
\label{eq:trace2a}
	\Tr\big\{\mathds{G}_{\rm r,1}^{\rm E,(2)}\big\}=
	\int\! \frac{\rd^2 q}{ 4 \pi^2}\,
	\re^{\ri \vq \cdot \vrh}\F{2}(-\vq)a_2(\vq), 
\end{align}
with
\begin{equation}
a_2(\vq)= \int \! \rd^2 \kappa'\,
	\vec{S}^{(2)}_1(\vq+\vk',\vk') \cdot 
	\vec{A}^{\rm (E)}_{\rm tr}(\vq+\vk',\vk') \;,
\end{equation}
and is evaluated in the same manner. The other contribution contains
a further integral,
\begin{align}
\label{eq:trace2b}
	\Tr\big\{\mathds{G}_{\rm r,2}^{\rm E,(2)}\big\} & = 
	\int\! \frac{\rd^2 q}{4 \pi^2} \,
	\re^{\ri \vq \cdot \vrh} \F{1}(-\vq)
	\int\! \frac{\rd^2 q'}{4 \pi^2}\\
	& \times \, \re^{\ri \vq' \cdot \vrh} \F{1}(-\vq') a_3(\vq,\vq') \, ,
\nonumber  
\end{align}
with integration variables $\vq=\vk-\vk'$ and $\vq'=\vk'-\vk''$, 
and with the expression 
\begin{align}
\label{eq:S2}
	a_3(\vq,\vq')& =\int\! \rd^2\kappa'' \,
	\re^{\ri(k_{z0}+k_{z0}'')z} \\
	& \times\vec{S}^{(2)}_2(\vk,\vk',\vk'')\cdot
	\vec{A}^{\rm (E)}_{\rm tr}(\vk,\vk'') \; ; 
\nonumber
\end{align}
again, $\vec{S}^{(2)}_1$ and $\vec{S}^{(2)}_2$ are stated explicitly in 
Appendix~\ref{App:A}. After computing $a_3(\vq,\vq')$ on a discrete mesh of 
$\vq$ and $\vq'$, a four-dimensional inverse FFT is performed for determining 
the resulting contribution to the trace of the Green's dyadic.

The integrals are numerically evaluated using Cuba routines;~\cite{Cuba} 
the Fourier transforms are executed by means of the FFTW library.~\cite{FFTW}

\section{Results}
\label{Sec:S_3}

In this section we discuss some numerical results for the local density 
of states above example topographies, calculated to second order in the 
profile height. We first consider the relative magnitude of the individual 
contributions, in order to estimate under which conditions the termination 
of the perturbation series can be justified. We then use the second-order
data to discuss the resolution power of an idealized NSThM in a mode of
operation in which the total heat transfer is kept constant while scanning
a sample's surface.

\subsection{Magnitude of second-order contributions}

\begin{figure}[tb]
\epsfig{file=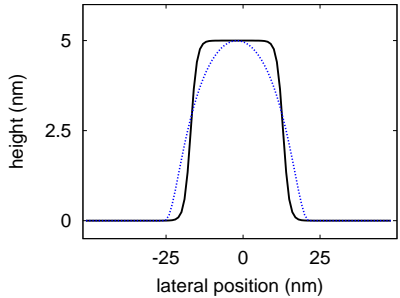, width=0.4\textwidth}
\caption{(Color online) Profile function~(\ref{eq:fermi}) with parameters 
	as employed in our model calculations, height $h=5\, \text{nm}$, 
	width $w=30 \, \text{nm}$, and inverse smoothing length 
	$\zeta=10^9 \, \text{m}^{-1}$ (full line), together with the
	reference profile~(\ref{eq:rund}) (dashed). In either case, the 
	dielectric properties of the sample are given by the Drude model 
	with parameters for gold at $300 \, \text{K}$.}   
\label{Fig:F_1}
\end{figure}

Our basic model structure is a bar with height~$h$ and smoothed edges placed
on an otherwise perfectly planar surface, infinitely extended in $y$-direction 
and possessing the width~$w$ in $x$-direction, as described by the function
\begin{equation}
\label{eq:fermi}
	h f_1(x)=h \frac{1}{\exp\big[\zeta(|x|-0.5w) \big]+1} \; .
\end{equation}
Our calculations are done for $h=5\, \text{nm}$, $w=30 \, \text{nm}$, and 
inverse smoothing length $\zeta=10^9 \, \text{m}^{-1}$. For comparison, we 
also consider the somewhat rounder profile
\begin{equation}
\label{eq:rund}
	h f_1(x) = h \exp\!\left(-\frac{1}{1 - (x/v)^2} +1 \right) \; , 
\end{equation}
with~$v$ adjusted such that the respective areas under the two 
functions~(\ref{eq:fermi}) and~(\ref{eq:rund}) coincide. The resulting 
profile shapes are drawn in Fig.~\ref{Fig:F_1}.

Because these profiles depend on only one variable, the Fourier transforms 
$\F{n}$ contain delta functions, so that the integrals over $q$ and $q'$ 
in Eqs.~\reff{eq:trace1}, \reff{eq:trace2a}, and \reff{eq:trace2b} become 
effectively one-dimensional, drastically reducing the numerical effort. 
The profiles are discretized with a stepsize of $1\, \text{nm}$, covering a 
total range of $500\, \text{nm}$; we have checked that the numerical results 
thus obtained are stable against further reduction of the grid size to 
$0.5\, \text{nm}$. We assume that the dielectric function $\epsilon(\omega)$ 
of the samples is given by the Drude model~\cite{AshcroftMermin76} 
\begin{equation}
	\varepsilon(\omega) = 
	1 - \frac{\omega_{\rm p}^2}{\omega^2 + \ri\gamma\omega}
\end{equation}
with plasma frequency $\omega_{\rm p} = 1.4 \times 10^{16} \, \text{s}^{-1}$ 
and inverse relaxation time $\gamma = 3.3 \times 10^{13} \, \text{s}^{-1}$, 
describing gold at the temperature $T = 300$~K.

\begin{figure}[bt]
\epsfig{file=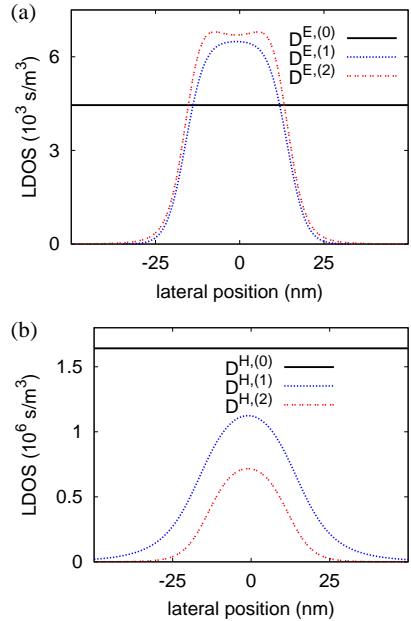, width=0.4\textwidth}
\caption{(Color online) Zeroth-, first-, and second-order contributions to the 
	electric~(a) and to the magnetic~(b) part of the LDOS above a gold 
	surface structured with the nanobar~(\ref{eq:fermi}) as sketched in 
	Fig.~\ref{Fig:F_1}, for $\omega =10^{14}\, \text{s}^{-1}$ at an 
	observation height of $10 \, \text{nm}$ above the base substrate 
	plane. Note the different scales.}
\label{Fig:F_2}
\end{figure}

In Fig.~\ref{Fig:F_2} we depict the zeroth-, first-, and second-order 
contributions to the electric and to the magnetic part of the LDOS for 
the structure~(\ref{eq:fermi}), evaluated at a constant height of 
$10 \, \text{nm}$ above the base plane for $\omega=10^{14}\, \text{s}^{-1}$, 
roughly equal to the dominant thermal frequency at $300 \, \text{K}$.
The second-order terms qualitatively show the same behavior as the first-order 
ones,~\cite{BiehsEtAl08} but there is a notable difference between the electric
and the magnetic part: The second-order contribution to the magnetic LDOS at
least is smaller than its first-order precessor, although only by a factor 
which is not small compared to unity, whereas the magnitude of the second-order 
electric contribution remains comparable to that of the first-order one, and 
even slightly exceeds it. At the bar's center, the first- and the second-order 
electric contributions amount to roughly 1.5 times the zero-order term. This 
behavior is not accidental; it can be quantitatively understood with the help 
of an elementary consideration: At sufficiently short distances, that is, 
for~$z-h$ not too large compared to the profile width~$w$, the local 
geometry equals that of a flat surface, shifted by~$h$ against the base plane. 
Therefore, the electric LDOS $D^{\rm E}$ at such a point $\vec{r}=[0,0,z]^t$ 
is approximately given by the LDOS $D^{\rm E}_{\rm fs}$ pertaining to a 
perfectly flat surface~\cite{JoulainEtAl03} through the relation
\begin{equation}
	D^{\rm{E}}(\vec{r}) \approx D^{\rm E}_{\rm fs}([0,0,z-h]^t) \; .
\label{eq:local}	
\end{equation}
Now the distance dependence of $D^{\rm E}_{\rm fs}$ in the near field is
given by
\begin{equation}
	D^{\rm E}_{\rm fs}([0,0,z]^t) \sim \frac1{z^3} \; ;
\label{eq:eldos}	
\end{equation}
the strong decay of this electric LDOS with the third power of the distance
clearly aids the local approximation~(\ref{eq:local}). Thus, one has
\begin{equation}
	D^{\rm{E}}([0,0,z]^t)\sim\frac1{(z-h)^3} \; .
\end{equation}
Expanding in powers of $h$, this yields 
\begin{equation}
	D^{\rm{E}}([0,0,z]^t) \sim \frac1{z^3} + \frac{3 h}{z^4}
	+ \frac{6 h^2}{z^5} + O(h^3) \; ,
\end{equation}
allowing one to estimate the ratios of the contributions appearing in
different orders:
\begin{equation}
	\frac{D^{\rm{E},(1)}}{D^{\rm{E},(0)}} \approx 3\frac{h}{z} \; ;
	\quad
	\frac{D^{\rm{E},(2)}}{D^{\rm{E},(0)}} \approx 6\frac{h^2}{z^2} \; . 
\end{equation}
With $h/z = 0.5$, which is the value considered in Fig.~\ref{Fig:F_2}, one 
obtains $D^{\rm{E},(1)} / D^{\rm{E},(0)} = D^{\rm{E},(2)} / D^{\rm{E},(0)}
= 1.5$, in quite good agreement with the exact numerical result. Generalizing
this argument to any order~$n$, one observes
\begin{equation}
	\frac{D^{{\rm E},(n+1)}}{D^{{\rm E},(n)}} \approx 
	\frac{n+3}{n+1} \, \frac{h}{z} \; .
\label{eq:este}	
\end{equation}
Thus, while the series may still converge for any $z > h$, convergence at
short distances would be rather slow; for $h/z = 0.5$ the magnitude of the
leading successive contributions, normalized to the zeroth-order one, is 
$1 : 3/2 : 3/2 : 5/4 : 15/16 : 21/32 : \ldots \; .$ In this example, 
terminating the perturbation series at the second order means that one 
collects only about half of the exact value. 

In the case of the magnetic LDOS, one has
\begin{equation}
	D^{\rm H}_{\rm fs}([0,0,z]^t) \sim \frac1{z} \; ,
\end{equation}
so that here a larger area of the sample's surface contributes to the LDOS than 
in the electric case~(\ref{eq:eldos}), implying that a local approximation
analogous to Eq.~(\ref{eq:local}) cannot be expected to work as well as before. 
Ignoring this restriction and performing the analysis nonetheless, one ends up 
with 
\begin{equation}
	\frac{D^{{\rm H},(n+1)}}{D^{{\rm H},(n)}} \approx 
	\frac{h}{z} \; ,
\label{eq:esth}	
\end{equation}
which, in view of the shaky foundation of the reasoning, still works
satisfactorily, capturing both the correct trends and the orders of magnitude
read off from Fig.~\ref{Fig:F_2}. 

\begin{figure}[tb]
\epsfig{file=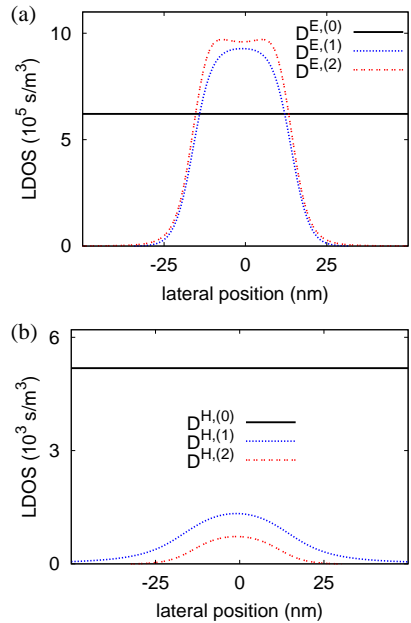, width=0.4\textwidth}
\caption{(Color online) As Fig.~\ref{Fig:F_2}, but for a sample consisting 
	of gallium nitride (GaN).}
\label{Fig:F_3}
\end{figure}

\begin{figure}[Hhbt]
\epsfig{file=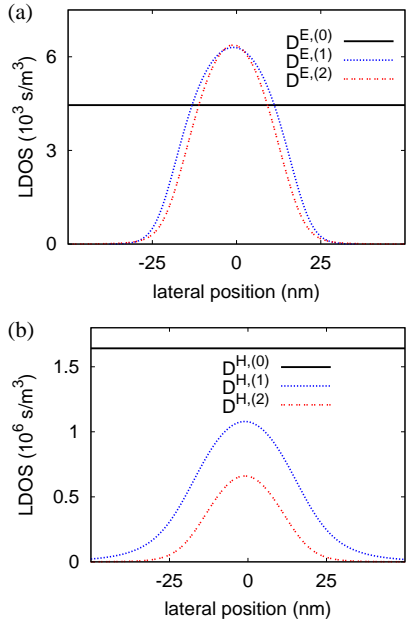, width=0.4\textwidth}
\caption{(Color online) As Fig.~\ref{Fig:F_2}, for a gold sample with the 
	reference profile~(\ref{eq:rund}).}
\label{Fig:F_4}
\end{figure}

It is evident that these general findings do {\em not\/} depend on the 
specific dielectric properties of the material. Indeed, when considering 
a polar sample with a permittivity described by the Reststrahlen 
formula~\cite{Klingshirn05}
\begin{equation}
	\epsilon(\omega) = \epsilon_{\infty}
	\left(1+\frac{\omega_{\rm L}^2 - \omega_{\rm T}^2}
	             {\omega_{\rm T}^2 - \omega^2 - \ri\gamma\omega}\right)
\end{equation}
and inserting parameters appropriate for gallium nitride~\cite{Adachi04}, 
namely, $\epsilon_{\infty} = 5.35$ for the high-frequency permittivity,
$\omega_{\rm L} = 1.41 \times 10^{14}\,\rm{s}^{-1}$ and
$\omega_{\rm T} = 1.06 \times 10^{14}\,\rm{s}^{-1}$ for the frequencies 
of the longitudinal and transversal phonons, and 
$\gamma=1.51 \times 10^{12}\,\rm{s}^{-1}$ for the relaxation rate, and
again taking the profile~(\ref{eq:fermi}), we obtain Fig.~\ref{Fig:F_3},
which shows the same qualitative features as the previous Fig.~\ref{Fig:F_2}
for the gold sample, although, of course, the scales are quite different;
now the electric contribution dominates. Likewise, the results do not seem 
to depend sensitively on the precise form of the structure: The corresponding 
data obtained for the reference profile~(\ref{eq:rund}) are remarkably similar 
to the previous ones, as shown in Fig.~\ref{Fig:F_4}, and again confirm the 
simple estimates (\ref{eq:este}) and (\ref{eq:esth}). For this reason, we only 
consider the gold nanobar~(\ref{eq:fermi}) in the following.

\begin{figure}[Hhbt]
\epsfig{file=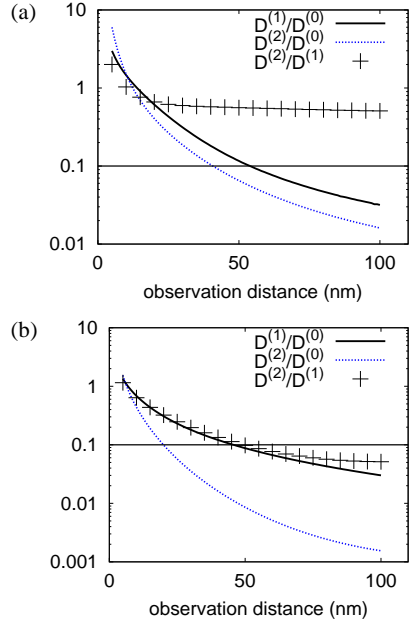, width=0.4\textwidth}
\caption{(Color online) Ratios of the zeroth-, \mbox{first-,} and second-order 
	contributions to the electric~(a) and to the magnetic~(b) LDOS, for 
	$\omega =10^{14}\, \text{s}^{-1}$ at an observation point $x = 0$
	above the center of the bar~(\ref{eq:fermi}).}    
\label{Fig:F_5} 
\end{figure}

To conclude the above discussion, in the cases studied so far the 
restriction to second-order perturbation theory already seems questionable, 
although the strong dominance of the magnetic LDOS for metallic samples might 
still mask the problem with the electric one. This is potentially important 
for NSThM-applications, where typical probe-sample distances range down to 
a few nanometers. On the other hand, low-order perturbation theory may be 
expected to work reliably when the profile height~$h$ clearly is the smallest 
length scale of the problem; according to the above reasoning, it should 
become better when increasing the observation distance~$z$. In order to 
estimate the smallest~$z$ at which second-order perturbation theory might 
still give quantitatively good results for our model profile~(\ref{eq:fermi}), 
we plot in Fig.~\ref{Fig:F_5} the ratios of the various contributions, 
evaluated above the bar's center at varying distance; here we also consider 
the regime $z - h \gg w$ where the short-distance estimates (\ref{eq:este}) 
and (\ref{eq:esth}) may no longer be taken for granted. As a rough guideline, 
one may accept the truncation of the perturbation series at the second order 
if the ratio of the second-order contribution to the zeroth-order one drops 
below $10\%$, say. For the dominant magnetic part this criterion is satisfied 
for $z > 20 \, \text{nm}$, while the electric part then requires 
$z > 41 \, \text{nm}$. (When reducing the acceptance limit to $5\%$, one 
gets $z > 26 \, \text{nm}$ and $z > 56 \, \text{nm}$, respectively.) The 
potential problem of slow convergence here is expressed by the fact that the 
ratio of second- to first-order contribution decreases only rather slowly with 
increasing~$z$. Still, in view of the relative smallness of the electric LDOS 
above metallic samples this does not seem to be essential.

\subsection{Resolution of an NSThM}

Scanning the LDOS at a constant height above the base plane, as done 
numerically in Figs.~\ref{Fig:F_2}, \ref{Fig:F_3}, and \ref{Fig:F_4}, 
corresponds to the {\it constant height} 
mode of operation of an NSThM. This is not an advantageous mode, for two 
practical reasons: On the one hand, it is hard to realize with sufficient
accuracy, on the other, it contains the risk of a probe-sample collision
when scanning a surface with an unknown topography, almost always resulting 
in irreparable damage to the delicate sensor.~\cite{WischnathEtAl08} A much 
more favorable mode avoiding these complications is the {\it constant transfer}
mode, meaning that the sensor height is continuously regulated such that the 
detected heat current remains constant when moving the sensor over the 
surface; the information about the sample's near field then is embodied in 
the recorded sensor height. Note that this latter mode differs from the 
{\it constant distance} mode, which uses additional information on the local 
distance of the sensor from the structured surface (obtained by electron 
tunneling spectroscopy) in order to keep that distance constant. 
That constant distance mode was employed experimentally by Kittel 
{\it et al.};~\cite{KittelEtAl08} numerical first-order results pertaining
to this mode can be found in Ref.~\onlinecite{BiehsEtAl08}. In contrast, 
a major benefit of the constant transfer mode lies in the fact that it 
exclusively requires heat-transfer information, so that there is no need 
to retain the sensor's ordinary STM-capability. Moreover, it allows one to 
assess the resolution limit of the NSThM in a simple manner.

For modeling this constant transfer mode, we select some appropriate fixed 
value of the LDOS, and then calculate that observation distance~$a$ at which 
this LDOS-value is reached. Only the sum of all contributions is of interest
now. For consistency, we also require that the second- to zeroth-order ratio 
remains less than $10\%$ for the magnetic LDOS, as the electric one does not 
contribute significantly here. Discussing the NSThM's resolution power first 
requires the specification of a norm structure containing the length scale~$s$ 
to be resolved. Here we take two parallel gold bars of the 
form~(\ref{eq:fermi}), as described by the profile  
\begin{align}
\label{eq:2bars}
	hf_2(x) = & 
	h\Bigg[\frac{1}{\exp\big[\zeta_1(|x+0.5s|-0.5w_1) \big]+1} \\
	& +\frac{1}{\exp\big[\zeta_2(|x-0.5s|-0.5w_2) \big]+1}\Bigg] \; ; 
\nonumber
\end{align}
the length scale in question is their separation~$s$. For our 
matter-of-principle calculations we again choose $h=5\,\text{nm}$, together 
with $w_1=w_2=30 \, \text{nm}$ and $\zeta_1=\zeta_2=10^9\,\text{m}^{-1}$.

\begin{figure}[tb]
\epsfig{file=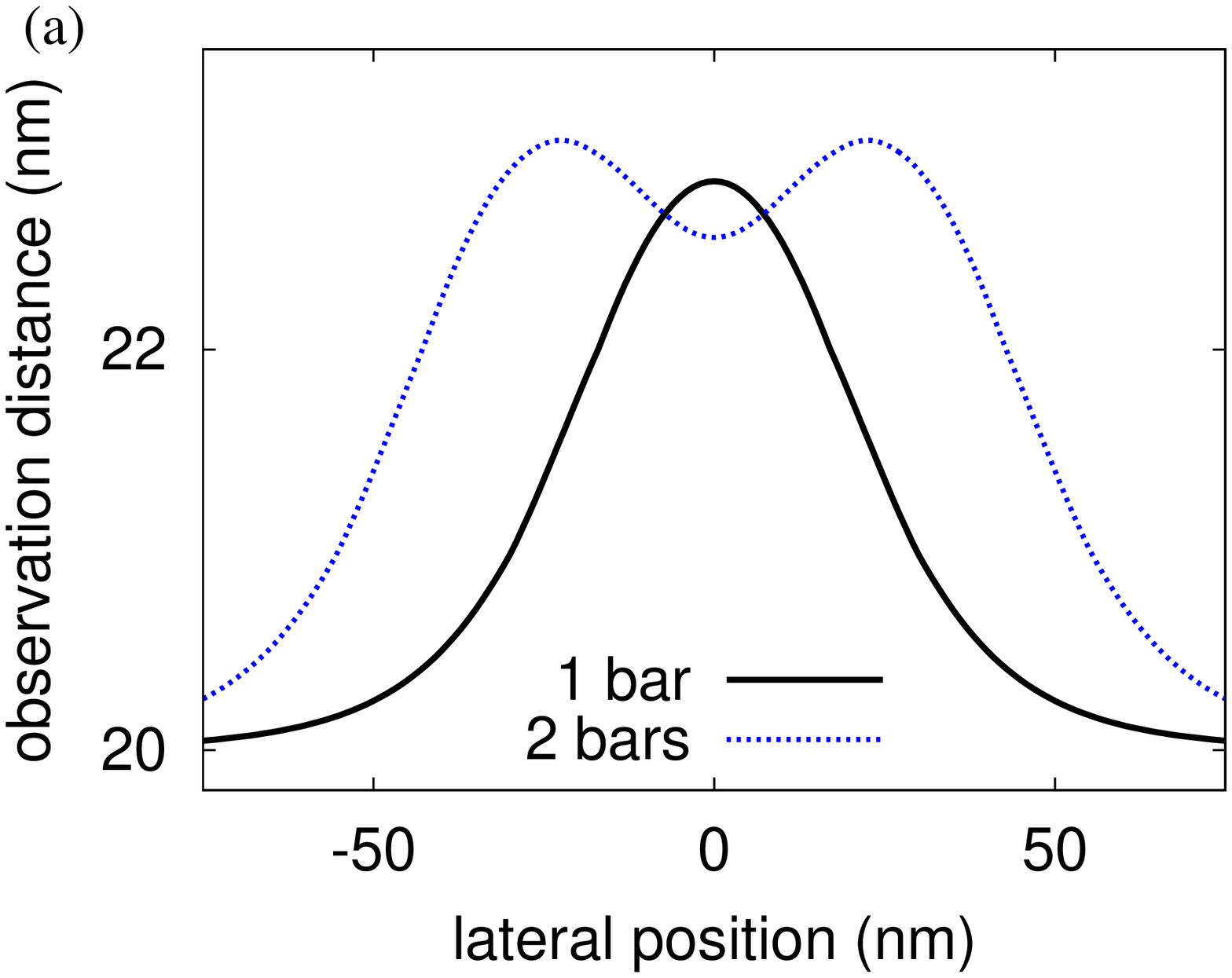, width=0.4\textwidth}
\epsfig{file=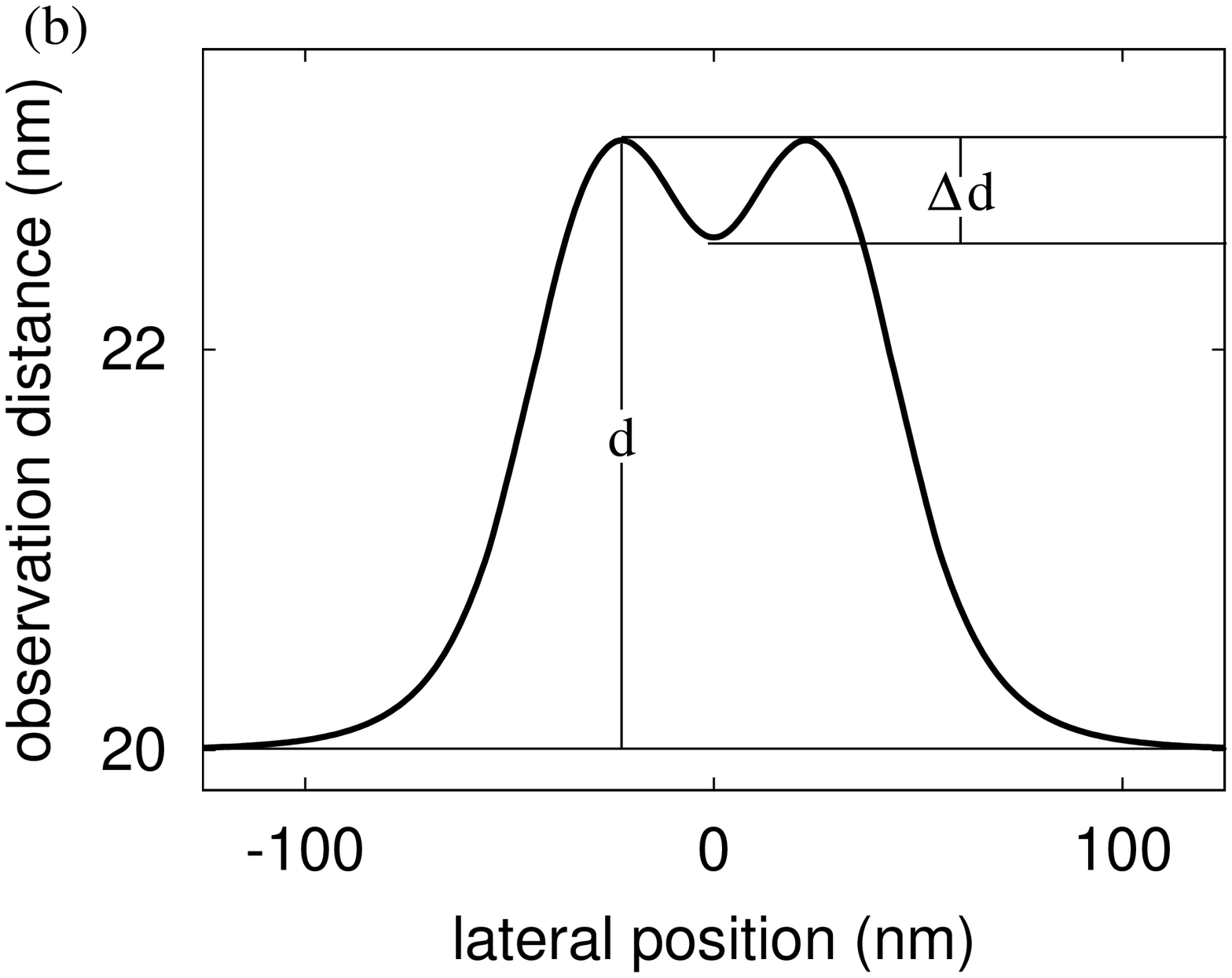, width=0.4\textwidth}
\caption{(Color online) (a) Observation distance above the one-bar gold sample 
	described by Eq.~\reff{eq:fermi} (1 bar), and above the sample with 
	two parallel bars specified by Eq.~\reff{eq:2bars} (2 bars), determined
	such that the second-order LDOS for $\omega=10^{14}\, \text{s}^{-1}$ 
	constantly keeps that value which is attained for the distance 
	$a_{\rm eff}=20\,\text{nm}$ far away from the bars.
	(b) Definition of the quantities $d$ and $\Delta d$ employed for 
	discussing the resolution of an ideal NSThM.}
\label{Fig:F_6}
\end{figure}

In Fig.~\ref{Fig:F_6}(a) we display second-order results for both the 
one-bar geometry, and for the two-bar structure with bar separation  
$s = 50\,\text{nm}$. Here the observation distance~$a$ is computed such 
that the LDOS remains fixed at the value attained for the distance  
$a_{\rm eff} = 20 \, \text{nm}$ above the base plane at positions far away 
from the bars, always assuming $\omega=10^{14}\, \text{s}^{-1}$.

In order to discuss the resolution of an idealized NSThM, we make two further 
assumptions: First, we propose that the sensor is point-like, so that no 
effects due to the real sensor's extension are considered, implying that 
we aim at the sensor-independent best possible resolution limit. In reality,
the finite sensor size will lead to a lower resolution. Second, we assume 
that the signal recorded by the device is proportional to the LDOS at the 
dominant thermal frequency, which actually appears to be quite a good 
approximation for metallic samples.~\cite{KittelEtAl08} We then take the 
ratio $\Delta d/d$, where $d$ is the maximum difference $a-a_{\rm eff}$
encountered above each of the two identical bars, and $\Delta d$ denotes the 
difference between that maximum distance and the minimum distance adopted 
between the bars, as illustrated in Fig.~\ref{Fig:F_6}(b). We now stipulate
that the two bars can be resolved if $\Delta d /d \geq r$; this number~$r$ 
characterizes the sensitivity of the respective experimental set-up.

\begin{figure}[tb]
\epsfig{file=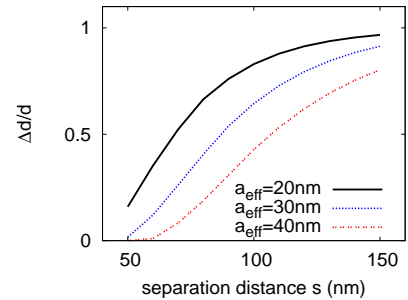, width=0.4\textwidth}
\caption{(Color online) Ratio $\Delta d/d$ for three different values of 
	$a_{\rm eff}$ ($20 \, \text{nm}$, $30 \, \text{nm}$, and 
	$40 \, \text{nm}$), as functions of the bar separation~$s$.}
\label{Fig:F_7}
\end{figure}

In Fig.~\ref{Fig:F_7} we plot the ratio $\Delta d/d$ for effective distances
$a_{\rm eff}=20 \, \text{nm}$, $30 \, \text{nm}$, and $40 \, \text{nm}$,
as functions of the bar separation~$s$. If we take $r = 0.2$ for the sake
of discussion, the resolvable minimum distances are  
$s_{\min}\approx 50 \, \text{nm}$ for $a_{\rm eff}=20 \, \text{nm}$, 
$s_{\min}\approx 65 \, \text{nm}$ for $a_{\rm eff}=30 \, \text{nm}$, and 
$s_{\min}\approx 80 \, \text{nm}$ for $a_{\rm eff}=40 \, \text{nm}$.
We emphasize that these figures serve to illustrate the basic principle
and should not be taken at face value. The resolution achievable with an 
actual NSThM device will also depend on the type of surface structure under 
investigation; a further limit will be imposed by the finite sensor volume. 
The key message, however, stands out clearly: When scanning an isothermal, 
nanostructured surface with a Near-Field Scanning Thermal Microscope, 
one is able to resolve structures with linear extensions which may fall 
orders of magnitude below the scale set by the dominant thermal wave 
length.~\cite{KittelEtAl08}

\section{Conclusions}
\label{Sec:S_4}

We have demonstrated that a numerical evaluation of Greffet's 
perturbation series for the scattered electromagnetic field at rough 
surfaces~\cite{Greffet88} is routinely feasible up to second order. This 
allows one to evaluate the local density of states above surfaces with 
arbitrary profiles, thus lifting the restriction to the limited class of 
profiles which can be dealt with analytically.

The convergence properties of this series seem to warrant further analysis.
While one may reasonably guess that low-order perturbation theory should be
sufficient when the profile height~$h$ is by far the smallest length scale 
of the problem,  the slow decrease of the successive contributions to the 
electric LDOS depicted in Fig.~\ref{Fig:F_5}(a) for the smoothed, but still
quite steep metallic model structure sketched in Fig.~\ref{Fig:F_1}, together
with the elementary estimates based on Eq.~(\ref{eq:local}), are warning signs.
While our results have been obtained for specific model profiles, it would be 
quite helpful to have mathematically rigorous and sharp upper bounds on the 
higher-order contributions for any given surface structure. 

We did not discuss possible effects due to the nonlocal dielectric response
of the sample, which might come into play at distances of a few 
nanometers.~\cite{HenkelJoulain06} The question whether such effects would
be detectable with a Near-Field Scanning Thermal Microscope (NSThM)
deserves further investigations.   
   
With respect to NSThM surface imaging, we have shown how to estimate the 
best possible resolution limit, attained for a point-like sensor. Here we 
have introduced a mode of operation characterized by constant heat transfer, 
giving access to isolines of the LDOS. It is now a major task to extend the 
preliminary studies reported in Ref.~\onlinecite{KittelEtAl08}, considering 
surfaces with both regular and random nanostructures, and to further explore 
the concept of near-field thermal imaging.

\begin{acknowledgments}
We wish to thank Achim Kittel, Uli F.\ Wisch\-nath, David Hellmann, 
Lars Hoelzel, Ludwig Worbes, and J\"urgen Parisi for continuing 
discussions of their experiments. 
This work was supported by the Deutsche Forschungs\-gemeinschaft through 
Grant No.\ KI 438/8-1. 
Computer power was obtained from the GOLEM~I cluster of the Universit\"at 
Oldenburg. 
S.-A.~B.\ gratefully acknowledges support from the Deutsche Akademie der
Naturforscher Leopoldina under Grant No.\ LPDS 2009-7.
\end{acknowledgments}

\begin{appendix}

\begin{widetext}
\section{Calculational details}
\label{App:A}

In this Appendix we state the precise forms of the expressions which have 
been used in Subsection~\ref{Subsec:IIB}. The vector $\vec{S}^{(1)}$ 
introduced in Eq.~(\ref{eq:S1}), required for computing the first-order 
electric contribution~(\ref{eq:trace1}), contains the products of the 
transmission coefficients $t_{\rm s}(\vk)=(2 k_{z0})(k_{z0}+k_z)^{-1}$ and 
$t_{\rm p}(\vk)=(2 \sqrt{\epsilon} k_{z0})(\epsilon k_{z0}+k_z)^{-1}$, together 
with a convenient prefactor:
\begin{equation}
	\vec{S}^{(1)}(\vk,\vk') =
	\frac{k_0^2}{\kappa \kappa'}\frac{\epsilon-1}{16 \pi^2}
	\frac{\re^{-\ri(k_{z0}+k_{z0}')z}}{k_{z0}k_{z0}'}
	\left[ \begin{array}{c} 
	t_{\rm s}(\vk) t_{\rm s}(\vk') \vk \cdot \vk' \\
	-t_{\rm s}(\vk) t_{\rm p}(\vk') \frac{k_z'}{n k_0}(\vk \times \vk')_z \\
	-t_{\rm p}(\vk)t_{\rm s}(\vk')\frac{k_z}{n k_0}(\vk \times \vk')_z \\
	\frac{t_{\rm p}(\vk)t_{\rm p}(\vk')}{n^2 k_0^2}
	[n^2 \kappa^2 \kappa'^2-k_z k_z' (\vk \cdot \vk')]
       \end{array} \right] \; .
\end{equation}
The other vector $\vec{A}^{\rm (E)}_{\rm tr}$ appearing in Eq.~(\ref{eq:S1})
contains the traces of the dyadic products,
\begin{equation}
	\vec{A}^{\rm (E)}_{\rm tr}(\vk, \vk') = \frac{1}{\kappa \kappa'}
	\left[ \begin{array}{c}
	\vk \cdot \vk'\\
	-\frac{k_{z0}'}{k_0}\kp \\	
	-\frac{k_{z0}}{k_0}\kp\\
	\frac{1}{k_0^2}(\kappa^2\kappa'^2-k_{z0}k_{z0}'\skp)
       \end{array} \right] \; .
\end{equation}
For computing the magnetic contribution, this vector has to be replaced by
\begin{equation}
	\vec{A}^{\rm (H)}_{\rm tr}(\vk, \vk') = \frac{1}{\kappa \kappa'}
	\left[ \begin{array}{c}
	\frac{1}{k_0^2}(\kappa^2\kappa'^2-k_{z0}k_{z0}'\skp)\\
	\frac{k_{z0}}{k_0}\kp \\
	\frac{k_{z0}'}{k_0}\kp\\
	\vk \cdot \vk'
       \end{array} \right] \; .
\end{equation}
The vector $\vec{S}^{(2)}_1(\vk,\vk')$ determining the second-order 
term~(\ref{eq:trace2a}) is given by 
\begin{equation}
	\vec{S}^{(2)}_1(\vk,\vk') = 
	\ri \frac{k_0^2(\epsilon-1)}{16 \pi^2 \kappa \kappa'}
	\frac{\re^{-\ri(k_{z0}+k_{z0}')z}}{k_{z0}k_{z0}'} 
	\left[ \begin{array}{c}
	\ts \tss (k_z + k_z') \skp\\
	\\
	-\ts \tps \frac{k_z'}{n k_0}(k_z+k_z')\kp\\
	\\
	-\tp \tss \frac{1}{nk_0}(k_z k_z' +n^2 k_{z0}^2)\kp\\
	\\
	\tp \tps \frac{1}{n^2 k_0^2}\bigg[ \kappa^2 \kappa'^2 (n^2 k_z'+k_z)\\
	- \skp k_z' (k_zk_z' + n^2 k_{z0}^2) \bigg]
       	\end{array} \right] \; ,
\end{equation}
whereas the vector $\vec{S}^{(2)}_2(\vk,\vk',\vk'')$ entering into the 
expression~(\ref{eq:S2}), and thus into the other second-order 
contribution~(\ref{eq:trace2b}), takes the form
\begin{equation}
	\vec{S}^{(2)}_2(\vk,\vk',\vk'') = 
	-\ri  \frac{k_0^2 (\epsilon-1)\re^{-\ri(k_{z0}+k_{z0}'')z}}{8 \pi^2\kappa \kappa'^2 \kappa''} 
	\frac{k_z'-k_{z0}'}{k_{z0}''k_{z0}} 
	\left[ \begin{array}{c} 
	\ts \tsss \bigg[\skp \skps \\ 
	-\frac{k_z' k_{z0}'}{\kappa'^2+k_z' k_{z0}'}\kp \kps \bigg]\\
 	\\
 	-\ts \tpss \bigg[\frac{k_z''}{n k_0} \skp \kps \\
 	+ \frac{k_z'}{n k_0} \frac{\kappa'\kappa''+k_{z0}'k_z''\skps}
	{\kappa'^2+k_{z0}'k_z'}\kp \bigg]\\
 	\\
 	-\tp \tsss \bigg[ \frac{k_z}{n k_0} \kp \skps\\
 	+ \frac{k_{z0}' \kps}{n k_0} 
	\frac{n^2 \kappa^2 \kappa'^2-k_z k_z'\skp}{\kappa'^2+k_{z0}'k_z'} 
	\bigg] \\
 	\\
 	\tp \tpss \bigg[ \frac{k_z'' k_z}{n^2 k_0^2}\kp \kps\\
 	+\frac{1}{n^2 k_0^2} \frac{\kappa'^2 \kappa''^2 + k_{z0}'k_z''\skps}
	{\kappa'^2+k_{z0}'k_z'}\big[ n^2 \kappa^2 \kappa'^2 - k_z k_z' \skp 
	\big] \bigg] \end{array} \right] \; .
\end{equation}
\end{widetext}

\end{appendix}

\end{document}